%% file: main.tex
\documentclass[runningheads]{llncs}

\input{assets/packages.tex}	% Input the required packages
\input{assets/crnCommands.tex} % Input macro commands for CRNs

\begin{document}

% DOCUMENT TITLE AND AUTHOR INFORMATION
\input{assets/meta.tex}

% INPUT THE MAIN DOCUMENT SECTIONS
\input{sections/introduction}
\input{sections/problemStatement}
\input{sections/relatedWork}

\input{sections/prefixTree}

\input{sections/preProcessing}
\input{sections/conclusion}

\newpage
\bibliographystyle{template/splncs04}
\bibliography{assets/stateRep}

\end{document}

%% file: assets/packages.tex
\usepackage[T1]{fontenc} 
\usepackage{graphicx} 

\usepackage[breaklinks=true]{hyperref}
\usepackage[dvipsnames]{xcolor}

\usepackage{multirow} % Tabular cells spanning multiple rows
\usepackage{amsmath,amssymb,amsfonts} % AMS font packages
\usepackage{proof}
\usepackage{enumitem} 	% Customize lists
\usepackage{cleveref} 	% Clever references
\usepackage{tabularx} 	% Better tables
\usepackage{mathpartir}	% Provides mathpar environment
\usepackage{algorithm}	% Provides algorithm support
\usepackage[noend]{algpseudocode} % Removes endif from algorithms
\usepackage{booktabs}	% More table formatting
\usepackage{listings}	% Support for code listings
\usepackage{xargs}  	% Use more than one optional parameter in new commands
\usepackage{tikz}		% Fancy diagram support
\usetikzlibrary{arrows,automata,trees} % More Tikz
\usepackage{wrapfig}
\usepackage{xspace}

%% file: assets/crnCommands.tex
\newcommand{\st}{\ensuremath{\vec{s}}\xspace}
\newcommand{\state}[1]{\ensuremath{\st_{#1}}\xspace}  %% state
\newcommand{\nextSt}{\ensuremath{\st^{\prime}}\xspace} % next state
\newcommand{\stSet}{\ensuremath{\mathbf{S}}\xspace} %% state set
\newcommand{\initSt}{\ensuremath{\state{0}}\xspace} %% initial state
\newcommand{\exec}{\ensuremath{\varrho}\xspace}
\renewcommand{\path}[1]{\ensuremath{\rho(\exec{#1})}\xspace} %% path of an
\newcommand{\ctmc}{\ensuremath{\mathcal{C}}\xspace} %% CTMC
\newcommand{\ctmcFull}{\ensuremath{\ctmc = \langle \stSet, \initSt,
		\tranRateShort, \labelFunc \rangle}\xspace} %% CTMC tuple
\newcommand{\labelFunc}{\ensuremath{\mathbf{L}}\xspace} %% labeling function in CTMC

\newcommand{\tranRateShort}{\ensuremath{\mathbf{R}}\xspace} % Transition rate matrix
\newcommand{\tranRate}[2]{\ensuremath{\tranRateShort(#1, #2)}\xspace} % Transition 
\newcommand{\react}[1]{\ensuremath{\mathcal{R}_{#1}}\xspace} % reaction channel with a 
\newcommand{\reactSet}{\ensuremath{\mathfrak{R}}\xspace} % reaction set
\newcommand{\reactSetSize}{\ensuremath{\mathnormal{n}}\xspace} % n reactions
\newcommand{\reactSetFull}{\ensuremath{\reactSet = \{\react{1}, \dots, 
		\react{\reactSetSize}\}}\xspace} % set of reaction channels
\newcommand{\crn}{\ensuremath{\mathcal{M}}\xspace}
\newcommand{\crnFull}{\ensuremath{\crn = \langle \speciesSet,
		\reactSet, \initSt \rangle}\xspace} % Chemical Reaction Network (CRN)
\newcommand{\species}[1]{\ensuremath{\mathcal{X}_{#1}}\xspace} % species with a 
\newcommand{\speciesSet}{\ensuremath{\mathfrak{X}}\xspace}
\newcommand{\speciesSetSize}{\ensuremath{\mathnormal{m}}\xspace} % m species
\newcommand{\speciesSetFull}{\ensuremath{\speciesSet = \{\species{1}, \dots,
		\species{\speciesSetSize}\}}\xspace} % set of species

\newcommand{\reactionRateConst}[1]{\ensuremath{k_{#1}}\xspace} % reaction
\newcommand{\probOp}{\ensuremath{\mathsf{P}}\xspace} %% probabilistic
\newcommand{\ragtimer}{\textsc{Ragtimer}\xspace}
\newcommand{\prism}{\textsc{Prism}\xspace}
\newcommand{\storm}{Storm\xspace}
\newcommand{\stamina}{\textsc{Stamina}\xspace}
\newcommand{\modest}{\textsc{Modest}\xspace}

\newcommand{\yices}{Yices 2}

%% file: assets/meta.tex
\title{Prefix Trees Improve Memory Consumption in Large-Scale Continuous-Time Stochastic Models}
\titlerunning{Prefix Trees Improve Memory Consumption}

% List of authors, institutes, and orcidID
\author{
Landon Taylor\inst{1}\orcidID{0000-0002-4071-3625}
\and
Joshua Jeppson\inst{1}\orcidID{0000-0002-4177-7604} 
\and
Ahmed Irfan\inst{2}\orcidID{0000-0001-7791-9021}
\and
Lukas Buecherl\inst{1}\orcidID{0000-0002-4844-6605} 
\and 
Chris Myers\inst{3}
\and 
Zhen Zhang\inst{1}\orcidID{0000-0002-8269-9489}
}
\authorrunning{L.~Taylor et~al.}

\institute{
Utah State University, Logan, Utah, USA \and
SRI International, Menlo Park, California, USA \and
University of Colorado Boulder, Boulder, Colorado, USA
}

\maketitle

\begin{abstract}
	\input{sections/abstract.tex}
	\keywords{Vector Addition Systems, Probabilistic Model Checking, Explicit 
	State Representation, Data Structures}
\end{abstract}

%% file: sections/abstract.tex
Highly-concurrent system models with vast state spaces like Chemical Reaction Networks (CRNs) that model biological and chemical systems pose a formidable challenge to cutting-edge formal analysis tools. Although many symbolic approaches have been presented, transient probability analysis of CRNs, modeled as Continuous-Time Markov Chains (CTMCs), requires explicit state representation. For that purpose, current cutting-edge methods use hash maps, which boast constant average time complexity and linear memory complexity. However, hash maps often suffer from severe memory limitations on models with immense state spaces. To address this, we propose using prefix trees to store states for large, highly concurrent models (particularly CRNs) for memory savings. We present theoretical analyses and benchmarks demonstrating the favorability of prefix trees over hash maps for very large state spaces. Additionally, we propose using a Bounded Model Checking (BMC) pre-processing step to impose a variable ordering to further improve memory usage along with preliminary evaluations suggesting its effectiveness. We remark that while our work is motivated primarily by the challenges posed by CRNs, it is generalizable to all CTMC models.

%% file: sections/introduction.tex
\section{Introduction}
\label{sec:introduction}
Real-world continuous-time probabilistic systems, including \emph{Chemical Reaction Networks} (CRNs) and other stochastic \emph{Vector Addition Systems} (VASes), induce \emph{Continuous-Time Markov Chain} (CTMC) models. 
Formal analysis of CTMCs is a focal point in the study of safety-critical systems, particularly due to their relevance in the modeling of synthetic biological designs~\cite{Buecherl2021,Friedenberg2022,Madsen2015} as well as communication systems~\cite{Daws2004}, dynamic power management~\cite{Sesic2008}, autonomous vehicle control~\cite{Cizelj2011},
and other critical processes~\cite{Volk2018}.
The inherent complexity of CRNs arises from their often infinite or intractably large state spaces, which complicate both analysis and state storage~\cite{Buecherl2021}.
\emph{Probabilistic model checking} (PMC) can provide provable quantitative reliability guarantees for safety-critical systems, potentially revealing flaws and errors in the early stages of the design. However, models' large or infinite state spaces combined with a lack of advance model information can pose formidable challenges to state-of-the-art PMC tools.

Explicit state representation---storage of the true numerical value of each state variable in every state---is required to enable CTMC time-bounded transient numerical analysis. This means that symbolic state representations such as \emph{Binary Decision Diagrams} (BDDs) must first be converted to an explicit state representation before transient analysis can be performed~\cite{Hensel2021}.
Cutting-edge PMC tools including \storm~\cite{Storm} store explicit states in a hash map, which balances memory and time complexity. 
\noindent However, many seemingly simple CRN models still require prohibitive amounts of memory~\cite{Ragtimer,Stamina,Andriushchenko2024}.

\textit{We propose the novel application of a prefix tree data structure for explicit CTMC state storage to drastically improve the memory consumption of very large state spaces.}
By reducing duplication of stored data, prefix trees reduce the memory required by a very large state space while maintaining \emph{constant} lookup and insertion time. From a theoretical perspective, prefix trees offer equivalent complexity to hash maps with the potential to offer memory savings. The benchmark results presented in this paper show that a prefix tree is highly effective in reducing memory usage, \textit{often requiring about half the memory of a hash map}.
While this work is motivated by CRNs, it is applicable to \emph{all} CTMC models and is well-suited for all VASes.

Because the memory savings from a prefix tree depend on the ordering of variables, we further propose a pre-processing step based on \emph{Bounded Model Checking} (BMC)~\cite{bmc} that provides a heuristic to allow for additional memory savings, especially during guided state-space exploration. Preliminary evaluations support this approach.

\textit{Structure of the paper.}
Section~\ref{sec:problemStatement} describes challenges of improving memory usage for explicit state representation and presents our objectives.
Section~\ref{sec:relatedWork} gives background information and related work.
Section~\ref{sec:prefixTree} presents a prefix tree state representation method that overcomes the challenges, as demonstrated by benchmarking results.
Section~\ref{sec:preProcessing} describes a pre-processing method that uses BMC to improve memory savings during guided state-space exploration.
Section~\ref{sec:conclusion} concludes the work.

%% file: sections/problemStatement.tex
\section{Problem Statement}
\label{sec:problemStatement}

CTMC transient analysis requires explicit-state representation. In highly concurrent models, including CRNs and many VASes, this can require a prohibitive amount of memory, even in cutting-edge tools. 
\emph{Our primary objective is to reduce the memory required to store a large state space while preserving the time complexity and explicit state representation from existing methods.}

\subsection{Motivation}

A key challenge for analyzing CRNs is their tendency to exhibit infinite or intractably large state spaces. Due to a lack of knowledge at design time, 
variable bounds are often very conservative estimates to avoid excluding critical behavior. Thus, the state space of a CRN can include billions or trillions of states, which can easily exceed the memory capacity of the computers used to analyze these models.
The probability of CTMC time-bounded transient reachability is of significant interest for synthetic biological models, which often pose a formidable challenge to existing tools~\cite{Buecherl2021,Friedenberg2022,Cotner2021,caseStudiesRepo}. 
This necessitates specialized approaches, as existing data structures often fall short for in cutting-edge approaches~\cite{Buecherl2021,Andriushchenko2024} including state space truncation~\cite{Stamina} and guided state space construction~\cite{Ragtimer}. 

Our work is specifically tailored to address the unique challenges posed by large-scale CRNs, aiming to enhance the efficiency and effectiveness of their analysis and design. 
\textit{We present an explicit state representation approach applicable to a wide range of models, including CRNs and other VASes, as well as probabilistic and non-probabilistic models for which states can be represented as vectors.} We are motivated especially by the memory challenge provided by CRNs during CTMC transient analysis. Our data structure is specialized to improve memory usage for highly concurrent VASes.

\subsection{Objectives}

Overcoming the memory hurdle for CRN and VAS state storage poses several requirements. To improve memory consumption relative to the hash maps used in cutting-edge tools, we establish the following objectives:

\noindent\textbf{Objective 1: Reduce Memory Usage.}
To enable analysis of very large models, our method must consume less memory than cutting-edge tools, which currently store states in a hash map during state space exploration. This is the most pressing challenge, as existing tools are unable to effectively analyze some very large CRN models due to their large state spaces~\cite{Israelsen2023}. Achieving this objective involves retaining or improving the worst-case memory complexity of a hash map while showing reduced memory usage in empirical evaluation.

\noindent\textbf{Objective 2: Maintain Time Complexity.}
Hash maps have, on average, constant lookup and insertion time. Our method should preserve these attributes, though some time-memory tradeoff can be expected. Achieving this objective involves preserving the worst-case time complexity from a hash map.

\noindent\textbf{Objective 3: Retain Explicit State Space.}
It must remain possible to extract states from our representation to construct an explicit state space, maintaining the ability to perform CTMC transient analysis.

%% file: sections/relatedWork.tex
\section{Background and Related Work}
\label{sec:relatedWork}
Existing literature and tooling highlights the challenges associated with exploring these state spaces, especially in the context of rare events that occur with low probability. 
For example, \prism~\cite{Prism}, \storm~\cite{Storm}, \modest's \texttt{mcsta} tool~\cite{Hartmanns2014}, and \stamina~\cite{Stamina} are generic PMCs that employ various state-space reduction techniques. The \modest Toolset~\cite{Hartmanns2014} performs statistical model checking for rare events~\cite{Budde2018}, which generates probability approximations rather than performing the full CTMC transient analysis.

\subsection{Chemical Reaction Networks and Vector Addition Systems}
\label{sec:CRN}

\textit{Chemical Reaction Networks} (CRNs) are a language for modeling the dynamics of chemical reactions in biological and chemical processes. Formally, a CRN \crn\ is a tuple \crnFull containing \reactSetSize\ reactions \reactSetFull\ operating on \speciesSetSize\ variables (chemical species) \speciesSetFull such that a reaction \ensuremath{\react{i} : \alpha \species{1} \xrightarrow{\reactionRateConst{i}} \beta \species{2}} is read ``Reaction $i$ \textit{consumes} $\alpha$ molecules of \species{1} and \textit{produces} $\beta$ molecules of \species{2} with \textit{reaction rate constant} $\reactionRateConst{i}$''. The initial state \initSt is defined such that $|\initSt| = \speciesSetSize$ and $\forall 0 \leqslant i < \speciesSetSize \cdot 0 \leqslant \initSt[i] < \infty$.
CRNs generally follow the \emph{Stochastic Chemical Kinetic} (SCK) model assumption, which limits to two reactants per reaction and requires that each reaction occur nearly instantaneously~\cite{Myers2010}.

A CRN induces a CTMC in which state change (caused by a reaction) occurs in discrete amounts, but the probability of state change is governed by rates defined with respect to real-valued time. A CTMC \ctmc is a tuple \ctmcFull in which \stSet is a set of states representing the state space, \ensuremath{\initSt \in \stSet} is the initial state, \reactSetFull defines the transition rate matrix, and \ensuremath{\labelFunc : \stSet \rightarrow 2^{AP}} is a state labeling function with atomic propositions \ensuremath{AP}. The transition rate \tranRate{\st}{\nextSt} is determined by the propensity of \react{i}, assuming \react{i} is the only reaction that causes this state change. 
Reaction \react{i} is \textit{enabled} to occur in a particular state \st\ if it has a non-zero probability of occurring.

\emph{Vector Addition Systems} (VASes), equivalently Petri nets, describe transition systems that change state based on the addition of vectors. Formally, a VAS $\mathcal{V}$ is a tuple $\mathcal{V} = \langle m, \vec{s_0}, R \rangle$, where $m$ is the number of variables (i.e., the length of each vector), $\vec{s_0}$ is the initial state, and $R = \{ \vec{r_1}, \ldots \vec{r_n} \}$ is the set of transitions such that $\vec{s_0} \in \mathbb{Z}_{\geqslant 0}^m$ and $\forall i \in [1,n] \cdot r_i \in \mathbb{Z}^m$. 

A stochastic VAS introduces a rate (for continuous-time) or probability (for discrete-time) associated with each transition's execution.
A CRN induces a stochastic VAS because transitions are restricted to addition within a system of states that are representable as vectors~\cite{Ceska2019}. For our purposes, it is sufficient to represent states as vectors of nonnegative integers such that for any two states $\state{\alpha},\state{\beta}$, $|\state{\alpha}| = |\state{\beta}|$ and $\forall i \in [0,m] \cdot 0 \leqslant \state{\alpha}[i] < \infty \land 0 \leqslant \state{\beta}[i] < \infty$. While we have specialized our approach for CRNs, our approach is generalizable to all CTMCs.

\subsection{Real-life Chemical Reaction Networks}
\label{sec:motivating}
\label{sec:models}
Our motivating example is the \emph{modified yeast polarization}
model~\cite{Daigle2011}, which was adapted from the pheromone-induced G-protein
cycle in \textit{Saccharomyces cerevisia}~\cite{Drawert2010} with a constant ligand population that keeps it from reaching equilibrium~\cite{Roh2010}. It is a CRN with 8 chemical reactions over 7 species:
\[
\resizebox{\textwidth}{!}{
	$\displaystyle
	\begin{array}{lll}
		\react{1} : \ \emptyset \xrightarrow{0.0038} \textrm{R}, &
		\react{2} : \ \textrm{R} \xrightarrow{4.00\times 10^{-4}} \emptyset,~~~ 
		&
		\react{3} : \ \textrm{L} + \textrm{R} \xrightarrow{0.042} \textrm{RL} + 
		\textrm{L}, \\
		\react{4} : \ \textrm{RL} \xrightarrow{0.010} \textrm{R},~~~ &
		\react{5} : \ \textrm{RL} + \textrm{G} \xrightarrow{0.011} 
		\textrm{G}_\textrm{a} + \textrm{G}_{\textrm{bg}},~~~ &
		\react{6} : \ \textrm{G}_\textrm{a} \xrightarrow{0.100} 
		\textrm{G}_\textrm{d}, \\
		\react{7} : \ \textrm{G}_\textrm{d} + \textrm{G}_{\textrm{bg}} 
		\xrightarrow{1.05\times 10^{3}} \textrm{G},~~~ &
		\react{8} : \ \emptyset \xrightarrow{3.21} \textrm{RL}. & \\
	\end{array}$
}
\]
All reaction propensities are expressed in molecules per second. For species \ensuremath{[R, L, RL, G, G_{a}, G_{bg}, G_{d}]}, the initial state is
\ensuremath{\initSt = [50, 2, 0, 50, 0, 0, 0]}. This simple model faces significant memory issues because its species counts are theoretically unbounded. The CSL property of interest for transient analysis in this model is \ensuremath{\probOp_{=?}(\Diamond^{[0, 20]} \, \textrm{G}_\textrm{bg}=50)}, which asks: ``What is the probability that within 20 seconds, species $\textrm{G}_\textrm{bg}$ count reaches a value of $50$?'' This property is a rare event, as this behavior occurs with an extremely low probability, and it requires an extremely large state space for analysis~\cite{Ragtimer}. 
Thus, accurate probability calculation requires the enumeration of an extremely large set of states.

We also apply our data structure to multiple large and complex CRNs, specifically, genetic circuits. Genetic circuits are engineered DNA sequences designed to perform user-defined functions within a cell, comparable to the way digital electronic circuits operate in computers~\cite{Myers2010}. These circuits process biological inputs (e.g., the presence of a chemical or environmental signal) and generate outputs (e.g., fluorescent proteins, metabolic changes, or other cellular behaviors) through the regulation of gene expression.

Several models are based on the \textit{genetic circuit 0x8E}, originally introduced by Nielsen et al.~\cite{nielsen_genetic_2016}.
Each of the seven models based on this circuit includes 18 species and 15 reactions.
An additional variation of circuit 0x8E, first introduced in~\cite{Fontanarossa2020}, 
incorporates a different logic implementation of the same logic function, designed to mitigate the erroneous behavior observed in laboratory environments~\cite{nielsen_genetic_2016}. This model contains 19 species and 16 reactions.

Additional models are based on the genetic implementation of a \textit{Muller C-element}~\cite{Madsen2015}.
C-elements are critical state-holding gates commonly used in asynchronous circuit design to coordinate parallel processes. 
This circuit implements a speed-independent design, ensuring correct behavior regardless of how fast or slow the gates change states. Both a detailed model and an abstract model were used. The models each have 20 species and 18 reactions.

\subsection{Prefix Trees}

\textit{Prefix trees}, also known as tries, are graph-based data structures that have been widely used for natural language processing. Prefix trees are efficient for storing and retrieving strings, as they exploit common prefixes among stored data for fast lookups and compact storage. For state representation, we favor prefix trees over related alternatives (including R-trees, X-trees, and radix trees) due to their simplicity, explicit representation, and low time and memory overhead. They are commonly used to quickly search for and retrieve words, phrases, or other textual elements based on their prefixes~\cite{Fredkin1960,Knuth1998}. 

Prefix trees have also been used to compress and store sets of numerical data, showing improvement over hash maps in PTrie, a general-purpose prefix tree implementation~\cite{Jensen2017}. 
\textit{However, the application of prefix trees for the verification and analysis of stochastic systems has not been investigated.}
Existing state-of-the-art tools for CTMC analysis primarily use hash maps for state storage. We benchmark our prefix tree implementation against Storm~\cite{Storm}, which stores states using the 3-bit murmur hash\footnote{Storm's hash map functionality is implemented in open-source code at \newline {\scriptsize \url{https://github.com/moves-rwth/storm/blob/master/src/storm/storage/BitVectorHashMap.h}}}. 
Hash maps are popular because of their time and memory efficiency, but they face challenges in handling state spaces with high memory requirements. The memory complexity of the prefix tree does not exceed that of the hash map in the worst case, and it yields significant memory savings in the average and best cases. Results in Section~\ref{sec:ptbench} corroborate this.

Our prefix tree implementation is designed as a sparse tree; each node contains a small hash map that allows fast lookup of successors. While many prefix tree implementations are possible, we observe a reasonable time-vs-memory trade-off in this implementation.

\subsection{Binary Decision Diagrams}

BDDs~\cite{bdd} are an efficient \textit{symbolic} data structure for state representation. BDDs efficiently store and explore the state spaces of the models~\cite{Baier2008,Burch1992,Miner2004}. One of the key advantages of BDDs is their ability to exploit redundancy within Boolean functions, reducing memory usage compared to explicit state enumeration.

While BDDs have been successful in addressing the state-space explosion problem in various domains, their applicability is often limited by the sensitivity of their performance to variable ordering. Finding an optimal variable ordering can be challenging, and suboptimal orderings can lead to a significant increase in the size of the BDD, negating the benefits of the data structure~\cite{Raseen2008}.

To address the limitations of BDDs, researchers have proposed \emph{Multi-Terminal Binary Decision Diagrams} (MTBDDs)~\cite{deAlfaro2000,Fujita1988,Hermanns2003}. MTBDDs allow multiple terminal nodes, enabling the representation of functions with a range of values rather than Boolean values. This makes MTBDDs useful for modeling and analyzing systems with quantitative properties, such as transition rewards. 
Existing tools for CTMC analysis, such as \prism and \storm, have incorporated BDD-based techniques to represent and manipulate the state spaces of the models under study. Both tools construct BDDs by default, as CTMC steady-state analysis does not require explicit state representation. However, CTMC transient analysis relies on explicit state representation for probability calculations.

Prefix trees exploit CRNs' concurrency to reduce memory usage. Similar to BDDs, we find that variable ordering in the prefix tree has an effect on memory consumption. However, while BDDs store states \textit{symbolically} in a way that is optimized for Boolean manipulation, the proposed prefix tree maintains \textit{explicit} state storage, which enables full CTMC transient analysis.

%% file: sections/prefixTree.tex
\section{Prefix Trees for State Representation}
\label{sec:prefixTree}

A prefix tree meets the proposed objectives for efficient state representation. By leveraging concurrency to reduce data duplication, prefix trees achieve memory savings while maintaining time complexity and usefulness compared to hash maps. Although the proposed method is designed for CRNs, it is extensible to all CTMC models. Thus, this section uses the syntax and semantics of CRN and VAS described in Section~\ref{sec:models}. 
\subsection{Data Structure and Algorithms}

In natural language processing, prefix trees owe their efficiency to the fact that many words share prefixes. By storing each prefix only once, prefix trees avoid storing duplicate characters. Similarly, transitions in highly concurrent models---particularly in CRNs---only update a small subset of variables at each execution. Thus, given a suitable variable ordering, many states share a ``prefix'', and we avoid storing duplicates of these prefixes.

A prefix tree $\mathcal{T}$ is a directed acyclic graph with root node $\mathcal{T}.\textsc{root}$. The depth of the graph for a model with $m$ variables is $m+1$, or the root node plus one level for each variable. A strict order $<$ must be imposed on the model's species such that all nodes at depth $i$ correspond to values of \species{i}. The presence of an edge between a value for \species{i} and a value for \species{i+1} where $\species{i} < \species{i+1}$ indicates that there is a state in which the connected values of \species{i} and \species{i+1} are present. 

\begin{figure}
	\centering
	\includegraphics[width=0.55\textwidth]{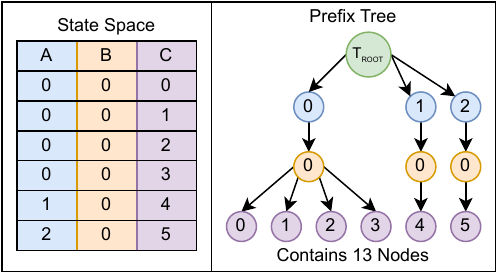}
	\caption{Example Construction of Prefix Tree from State Space}
	\label{fig:ss_to_trie}
\end{figure}

For example, in a model containing variables $\speciesSet = \{A,B,C\}$ such that $A<B<C$, a state $[2,0,5]$ is represented in the prefix tree as $\mathcal{T}.\textsc{root} \rightarrow 2 \rightarrow 0 \rightarrow 5$. This example is extended in Figure~\ref{fig:ss_to_trie}, which shows the state space (left) and the resulting prefix tree (right). In the figure, the state space is represented by one row per state, and the order imposed on the prefix tree is $A<B<C$.

% \ltzz{It looks like the extract states procedure in Algorithm~\ref{alg:prefixTree} needs to initialize the "current sequence" before calling the extract states recursive procedure (line 17) for the first time. The same initialization is missing for the "current node" on line 18.}
Algorithm~\ref{alg:prefixTree} formalizes the procedures for insertion, lookup, and state space extraction in the proposed data structure. Informally, the operations supported by this data structure are as follows:
\begin{description}
	\item[Insertion.] Add a new state to the tree by traversing the tree, following any existing edge corresponding to the ordered values of the state variables. Create nodes and edges after the first non-existent variable value.
	\item[Lookup.] Traverse the prefix tree, following the path corresponding to the ordered values of the state variables. If a complete path from root to leaf node exists, the state has been found.
	\item[State-Space Extraction.] Enumerate all the states stored in the prefix tree by performing a depth-first traversal of the structure.
\end{description}

In the case of Figure~\ref{fig:ss_to_trie}, insertion of the first state, $[0,0,0]$, requires the creation of three new nodes. However, the next state, $[0,0,1]$, requires that only one new node be created. The shared prefix $[0,0]$ is found by a graph traversal, then node $1$ is added at the lowest level of the prefix tree. The pattern continues for each state until the full prefix tree is constructed.

% \ltzz{The ExtractStates procedure in Algorithm~\ref{alg:prefixTree} needs some clarification. First, in line~\ref{alg:initCurSeq}, did you mean to initialize both the "current sequence" and $\mathcal{S}$ to an empty set? If so, do you need the curly braces? This also implies that the "current sequence" is a set. Should it be a list type instead, as it is a "sequence", which has an order? Second, line~\ref{alg:extractStatesRecur} is the declaration of the ExtractStatesRecursive procedure. I think it should be listed after the ExtractStates procedure as a standalone procedure. Then, the call to this recursive procedure inside ExtractStates should pass both the "current sequence" and $\mathcal{S}$ as parameters because ExtractStatesRecursive modifies both. Correct? Third, before line~\ref{alg:checkTerminal}, show how "current node" is obtained from ExtractStatesRecursive, which only has two in-parameters, but neither of them is the "current node". Lastly, after you factor ExtractStatesRecurive, double check the in-and out-parameters of each caller and callee to make sure that they match.}
% \zz{From Landon: I re-wrote the procedures to hopefully clarify this. Does the new version make more sense?}

\begin{algorithm}[tbh]
	\caption{Prefix Tree Operations}
	\label{alg:prefixTree}
	\begin{algorithmic}[1]
		\Statex \textbf{Data Structure:}
		\Statex \quad $\mathcal{T}$ = Prefix Tree
		\Statex \quad $\mathcal{T}.\textsc{root}$ = Root node of the prefix 
		tree
		\Statex \quad $\mathcal{T}.\textsc{isTerminal}(n)$ = Returns true if 
		node $n$ is a terminal node

		\Procedure{Insert}{$\mathcal{T}, s$}
		\Comment{$s$ is a sequence (state) to be inserted}
		\State $n \gets \mathcal{T}.\textsc{root}$
		\For{each $c \in s$}
		\If{$n$ does not have a child labeled $c$} \label{line:checkChild}
		\State $n.\textsc{AddChild}(c)$
		\EndIf
		\State $n \gets n.\textsc{GetChild}(c)$
		\EndFor
		\State $n.\textsc{MarkAsTerminal}()$
		\EndProcedure
		
		\Procedure{Search}{$\mathcal{T}, s$}
		\Comment{$s$ is a sequence (state) to be searched}
		\State $n \gets \mathcal{T}.\textsc{root}$
		\For{each $c \in s$}
		\If{$n$ does not have a child labeled $c$}
		\State \textbf{return} \textbf{false}
		\EndIf
		\State $n \gets n.\textsc{GetChild}(c)$
		\EndFor
		\State \textbf{return} $\mathcal{T}.\textsc{isTerminal}(n)$
		\EndProcedure
		
		\Procedure{ExtractStates}{$\mathcal{T}$}
		\Comment{Extract all stored sequences (states)}
		\State $\mathcal{S} \gets \emptyset$\label{alg:initCurSeq}
        \Comment{$\mathcal{S}$ is the resulting state space construction}
		% \State current sequence $\gets [\;]$
		% \State current node $\gets$ $\mathcal{T}.\textsc{root}$
		% \Comment{$\mathcal{S}$ is the resulting state space construction}
		\State $\mathcal{S} \gets \text{ExtractStatesRecursive}(
            \mathcal{T}.\textsc{root}, \;
            [ \; ], \;
            \mathcal{S}
		)$
		\State \Return $\mathcal{S}$
		\EndProcedure
        \Procedure{ExtractStatesRecursive}
        {%
            \text{node}, 
            \text{sequence},
            $\mathcal{S}$
        }\label{alg:extractStatesRecur}
        \If{$\mathcal{T}.\textsc{isTerminal}(\text{node})$} \label{alg:checkTerminal}
        \State $\mathcal{S} \gets \mathcal{S} \cup \{\text{
        sequence}\}$
        \EndIf
        \For{each child $c$ of node}
        \State $\mathcal{S} \gets \text{ExtractStatesRecursive}(c, \text{sequence} + c, \mathcal{S})$
        \EndFor
        \State \Return $\mathcal{S}$
		\EndProcedure
	\end{algorithmic}
\end{algorithm}

\subsection{Evaluation of Objective 1: Reduce Memory Usage}

In the worst case (i.e., no states share a common prefix), memory complexity is equivalent to a hash map: All values are stored exactly once. However, the prefix tree reduces the average memory load by grouping prefixes together and reducing the need to store an entire hash table.
% The memory efficiency of a prefix tree stems from its ability to share common prefixes among the stored states. This means that the total memory consumption of the tree is often significantly lower than the total memory required to store each state independently. 
Consider the example in Figure~\ref{fig:ss_to_trie}.
% , for example. 
A hash map would explicitly store the data in the table on the left, or effectively 18 values. Further, hash maps over-allocate memory to avoid collisions, so a time-efficient hash map may allocate memory for $\textit{next\_prime}(18\times 2)$, or 37, values. Because the prefix tree combines shared prefixes, it stores only 13 values.

\noindent\textbf{Memory Complexity of Prefix Trees.}
The \textsc{Insert} procedure in Algorithm~\ref{alg:prefixTree} is the only procedure that causes memory consumption. The main loop of the procedure iterates exactly $m$ times (where $m$ is the number of variables in the model), allowing up to a total of $m$ new nodes to be added to the prefix tree. 
If the condition on line~\ref{line:checkChild} is false, no new memory is consumed. 
% If it is true,
Otherwise, a new node is created and consumes memory. If the condition is \textit{always} true (i.e., the operation consumes the greatest amount of memory possible), a total of $m$ new nodes are added. Thus, adding one new state in the worst case requires $O(m)$ memory. Thus, inserting $n$ states has a worst-case memory complexity of $O(n \times m)$.
% The key advantage of prefix trees is their ability to exploit common prefixes among the stored states. 
In the best case, where all states share a prefix and only the final variable of each state is unique, the memory complexity of storing a single state can be significantly reduced to the best case: $\Omega(n \times 1)$. 
% \ltzz{Fix section reference below.}
In practice, the benchmarks presented in Section~\ref{sec:ptbench} demonstrate that average memory usage is somewhere between these two extremes.

The degree of memory savings achieved by the prefix tree depends on the structure and redundancy present in the state space. Models with a high degree of similarity among states, such as CRNs, are well-suited for prefix tree-based storage, as the structure exploits common patterns. 
% We further exploit this concurrent nature by exploiting variable bounds, as described in Section~\ref{sec:preProcessing}.

\noindent\textbf{Memory Complexity of Hash Maps.}
In contrast, the memory complexity of a hash map that stores identical states is fixed at $O(n \times m)$. This is because all $m$ values for all $n$ states are stored in their entirety in the hash map. In practice, a hash map has an additional large memory overhead, as hash tables over-allocate memory to prevent collisions and maintain constant average time complexity. A hash map stores the equivalent memory of one node per each variable, so its best-case memory consumption is also $\Omega(n \times m)$.

\noindent\textbf{Comparison.}
In the worst case, the memory complexity of both hash maps and prefix trees is $O(n \times m)$. However, the prefix tree's ability to avoid duplication allows its memory to decrease toward $\Omega(n \times 1)$, while the hash map's complexity is fixed at $\Omega(n \times m)$. In practice, benchmarks in Section~\ref{sec:ptbench} corroborate this, often showing a memory decrease of more than 50\% compared to a hash map.
% as seen in Section~\ref{sec:ptbench}. 
% \ltlb{You might want to add here: as seen in ....}

\subsection{Evaluation of Objective 2: Maintain Time Complexity}

Hash maps are often appealing because their lookup and insertion times are constant. In order to compete with hash maps, prefix trees must provide constant or near-constant time complexity.
% , allowing for a small time penalty in exchange for memory savings. 
The following analyses show the time complexity for a model with $m$ variables and $n$ states.

\noindent\textbf{Time Complexity of Prefix Tree Operations.}
The \textsc{Insert} and \textsc{Search} procedures each include one loop with exactly $m$ iterations, so the time complexity for each is $O(m)$. The \textsc{ExtractStates} procedure is a depth-first traversal with time complexity $O(m \times n)$, as it traverses up to $n$ paths of depth $m$. In most realistic models, we observe that $m \ll n$.

\noindent\textbf{Time Complexity of Hash Map Operations.}
Consider a model with $m$ variables and $n$ states. Insertion and lookup in a hash map with an efficient hashing algorithm each have an average time complexity of $\Theta(1)$, while the worst-case time complexity is $O(n)$~\cite{Cormen2009}. The extraction of state space requires the enumeration of exactly $n$ states, so its complexity is $\Omega(n)$.

\noindent\textbf{Comparison.}
A prefix tree and a hash map have comparable time complexity. The number of variables is constant (and often low) for each model. We thus suggest that the insertion and lookup complexities of a prefix tree remain competitive with those of a hash map, with a small time penalty in the case of the prefix tree.
% Intuitively, prefix tree operations are required to walk along a graph structure of depth $m$, where $m$ is the number of variables in the model. By comparison, hash maps require only one operation plus occasional re-hashing steps.
% \ltjj{Reviewer could ask why you're comparing $O$ to $\Theta$ complexity type (worst case vs average). I actually think if you mention that then it's a strength since you're saying the proposed method has at worst case linear complexity, where a hash map, while it's $\Theta(1)$ is $O(2\times \mathtt{nextprime}(n))$. Make sure to note that inherently $m << n$.}
While the prefix tree incurs a \textit{worst-case} time cost of $O(m)$ compared to the \textit{average-case} hash map time cost of $\Theta(1)$, the memory savings afforded by the prefix tree often justify the extra time cost, especially for very large CRN models, for which memory consumption is the primary limiting factor. 

\subsection{Evaluation of Objective 3: Retain Explicit State Space}
Because Algorithm~\ref{alg:prefixTree} presents a procedure (\textsc{ExtractStates}) to extract the explicit state space from the prefix tree, this method retains the usefulness of explicit state representation while affording memory savings. The prefix tree preserves the explicit state representation required for CTMC transient analysis. This explicit state representation is interpretable by a probabilistic model checker, maintaining compatibility with existing verification tools and workflows. 

% \subsection{Summary}

% %The use of prefix trees as a state storage mechanism has the potential to address the memory complexity challenges faced by current tools while retaining competitive time complexity and explicit state representation required for model analysis. 
% The procedures presented in Algorithm~\ref{alg:prefixTree} improve average memory consumption, maintain competitive time complexity, and afford equivalent utility compared to a hash map. Thus, prefix trees are a promising alternative to hash maps for models with very large, highly concurrent state spaces.

%Solution: Prefix Tree
%	Data structure description
%	    Algorithm pseudocode
%	    Nice figures
%	Challenge 1: less memory than hash map
%		Complexity analysis
%		Memory benchmarks
%		Show how PT reduces duplication
%		Hint at pre-processing
%	Challenge 2: Constant lookup/insert time
%		Complexity analysis
%		Time benchmarks
%	Challenge 3: State space construction
%		Show that it's comparable to a hash map

\subsection{Prefix Tree Implementation}
We implemented a custom prefix tree in native C++ linking against \texttt{glibc} for parity with \storm. It stores in each node a small hash map of species values to successors in order to mitigate large branching factors.
Two benchmarking programs execute identical state space exploration processes, storing and searching for states in either \storm's built-in hash map or the proposed prefix tree. To achieve this, we extended \storm's compressed \texttt{BitVector} class, which is used to store state variable values, to be indexable (i.e., we assign a unique index to each CTMC variable based on its location in the \texttt{BitVector} object). We use this index to override the C++ \texttt{[]} operator and iteration functionality for the \texttt{BitVector} class, which allows a state to be iteratively parsed for variable-by-variable lookup and insertion in the prefix tree.

The benchmarking programs perform breadth-first state-space exploration for a designated number of steps, timing the insertion and lookup operations, and recording memory usage. Benchmarks were performed on the models described in Section~\ref{sec:models} on a machine with an AMD Ryzen Threadripper 12-Core 3.5 GHz Processor and 132 GB of RAM, running Ubuntu Linux 22.04.5. The benchmarking programs and models are open source on Github\footnote{Benchmarking scripts and Storm interface are open-source at \\ \url{https://github.com/formal-verification-research/crn-prefix-tree}}.

Integration of the prefix tree into tools that are compatible with the Storm API is relatively straightforward. Such an integration can provide users with the option to reduce the memory requirements of checking models with vast state spaces. Integrating a prefix tree data structure for state storage in PMC tools may improve usability for vast, highly concurrent models.

\subsection{Prefix Tree Benchmarks}
\label{sec:ptbench}

We present benchmarks comparing Storm's built-in data structure to a custom prefix tree on variations on the realistic and challenging models with vast state spaces described in Section~\ref{sec:models}, performing stepwise state space exploration and construction for each model over $10^i$ steps, where $i \in [3,8]$. This range is used to balance the state space size with the need to benchmark a large set of models. We analyze the lookup time, the insertion time, and the total memory consumption for every combination of models and steps.

Table~\ref{tab:percent_improvement_by_tests} shows a summary of the improvement of the prefix tree over Storm's native hash map. Each row represents the average improvement seen for all models. The table shows the following data, respectively:

\begin{description}
    \item[Steps.] Number of steps explored. A step is the result of executing one transition.
    \item[Lookup.] Lookup time improvement of the prefix tree compared to hash map. 
    \item[Insert.] Insertion time improvement of the prefix tree compared to hash map.
    \item[Hash Map.] Memory consumed when using a hash map.
    \item[Prefix Tree.] Memory consumed when using a prefix tree.
    \item[Memory \%.] Memory improvement of the prefix tree compared to hash map expressed as percentage savings.
\end{description}

The time improvement values are negative because the prefix tree time complexity is slightly higher than the average hash map time complexity.
As the size of the state space (correlated to the number of steps) increases, the memory savings increase greatly. On average, the prefix tree's memory requirement is lower than the hash map's, and larger state spaces introduce more opportunities for shared prefixes. As expected, the prefix tree's lookup and insertion times are slightly higher than those of the hash map.
In models that require hours to analyze and severely constrain device memory, even the largest observed time penalty (about 15 minutes) is likely worth the memory savings it incurs (about 12 GB).
We observe that for large state spaces, the prefix tree data structure consistently consumes \emph{less than half the memory} of Storm's native hash map.

\begin{table}[tbh]
	\centering
	\setlength{\tabcolsep}{5pt}
	\caption{Average Improvement by Number of Steps}
	\label{tab:percent_improvement_by_tests}
	\begin{tabular}{llllll}
		\toprule
		Steps &     Lookup &      Insert &     Hash Map &  Prefix Tree & Improvement \\
		\midrule	
        $10^3$ &    0.00 s &     -0.01 s &     43.09 MB &     42.77 MB &      0.73\% \\
		$10^4$ &   -0.01 s &     -0.09 s &     45.79 MB &     42.91 MB &      6.29\% \\
		$10^5$ &   -0.12 s &     -0.93 s &     70.49 MB &     50.99 MB &     27.65\% \\
		$10^6$ &   -0.97 s &     -9.35 s &    291.59 MB &    135.65 MB &     53.48\% \\
		$10^7$ &   -6.63 s &    -94.79 s &      2.35 GB &      0.99 GB &     57.63\% \\
		$10^8$ &  -29.77 s &   -946.56 s &     21.61 GB &      9.58 GB &     55.67\% \\
		\bottomrule
	\end{tabular}
\end{table}

Table~\ref{tab:memory_savings} shows the 10 best performing models for memory savings (in bytes). We observe that highly concurrent models tend to provide better memory savings, as expected. In addition, memory savings increase as the state space grows, so the largest state spaces see the greatest savings. 

We also observe that in each of these models we are able to reduce memory consumption by \emph{over 50\%}; that is, for a reasonably small time trade-off, we see incredible memory savings. We argue that this data structure not only improves memory performance; it makes it \textit{feasible} to store some state spaces that were previously impossible to store. For example, a machine with 32 GB of memory would not be able to effectively store many of the models in Table~\ref{tab:memory_savings} with a hash map, but it would have no problem storing them with a prefix tree.
Further, constructing an entire state space for CTMC transient analysis on these models can require several hours to several days~\cite{Israelsen2023}, so we argue that time penalties of seconds to minutes are not excessive.

\begin{table}[hbt]
	\centering
	\setlength{\tabcolsep}{5pt}
	\caption{Top 10 Models with Greatest Memory Savings}
	\label{tab:memory_savings}
	\begin{tabular}{lllll}
		\toprule
		Model & Steps & Hash Map & Prefix Tree & Improvement \\
		\midrule
		Circuit0x8E\_000to011 & $10^8$ &                34.14 GB & 12.86 GB & 62.33\% \\
		Circuit0x8E\_000to101 & $10^8$ &                34.14 GB & 12.86 GB & 62.33\% \\
		Circuit0x8E\_LHF\_000to011 & $10^8$ &           35.40 GB & 14.23 GB & 59.80\% \\
		Circuit0x8E\_101to011 & $10^8$ &                33.97 GB & 12.89 GB & 62.06\% \\
		Circuit0x8E\_101to000 & $10^8$ &                33.97 GB & 12.89 GB & 62.06\% \\
		Circuit0x8E\_011to101 & $10^8$ &                33.94 GB & 12.88 GB & 62.08\% \\
		Circuit0x8E\_011to000 & $10^8$ &                33.94 GB & 12.88 GB & 62.08\% \\
		Speed\_Independent\_10\_10 & $10^8$ &           27.68 GB &  8.84 GB & 68.06\% \\
		SimplifiedMotilityRegulation & $10^8$ &         21.86 GB &  9.06 GB & 58.55\% \\
		Speed\_Independent & $10^8$ &                   19.68 GB &  8.75 GB & 55.54\% \\
		\bottomrule
	\end{tabular}
\end{table}

%% file: sections/preProcessing.tex
\section{Pre-Processing to Improve Memory Usage for Guided State Space Exploration}
\label{sec:preProcessing}

When PMC for large models requires an intractible amount of resources, guided state-space exploration is a compelling alternative. Existing tools include \ragtimer~\cite{Ragtimer}, which explores a state space from a set of reachable traces, and \stamina~\cite{Stamina}, which truncates a state space to reduce the model's memory load, depend on memory-efficient state storage methods. 

State-space exploration and truncation rely heavily on the ability to store and look up states on-the-fly. Because it is desirable to explore the largest possible subset of state space during guided exploration, it is useful to increase the number of states that can be stored in a finite amount of memory. Intuitively, a larger state space provides a better chance of obtaining accurate probability bounds. 
\emph{Therefore, decreasing the memory consumed by each state can enable the construction of larger state spaces in order to improve probability calculation.}

\paragraph{Exploiting Variable Ordering in Prefix Trees.}
One way to save memory is to exploit the ordering of variables in a prefix tree. For that, we first calculate finite-step bounds for each variable to predict a range of reachable values for each variable. Then, we use the bounds to generate an efficient variable ordering to impose for prefix tree construction.
Recent advances show that \emph{bounded model checking} (BMC) is an effective way to synthesize variable bounds in CRNs~\cite{Ahmadi2024}. We extend this by generating a \emph{range} of bounds. 

\paragraph{Transition System and Bounded Model Checking.}
A \textit{symbolic transition system} $\mathcal{S} = \langle X, I, T\rangle$ is a tuple where $X$ is a finite set of (state) variables, $I(X)$ is a logical formula that represents the initial states of the system, and $T(X, X')$ is a formula that describes its transition relation, with $X'$ being the set obtained by replacing each element $x$ in $X$ with $x'$. A state $s_i$ of $\mathcal{S}$ is an assignment to the variables in $X$.
A trace $\sigma_k = s_0, s_1, s_2, \ldots, s_{k-1}$ of length $k$ for $\mathcal{S}$ is a sequence of states where $s_0$ satisfies $I(X)$, and $s_i \land s_{i+1} \text{($X$ renamed to $X'$)}$ satisfies $T(X, X')$ for all $0 \leqslant i < k-2$. Let the set obtained by replacing $x$ with $x@i$ be denoted as $X@i$.
\emph{Bounded Model Checking} (BMC)~\cite{bmc,bmc2} is a symbolic model checking technique that examines all paths of the system from the initial state up to a fixed length, using a \emph{Satisfiability} (SAT)~\cite{satsolver} or \emph{Satisfiability Modulo Theories} (SMT)~\cite{smt} solver. 
For a transition system $S = \langle X, I, T\rangle$ and an invariant property $P(X)$, i.e. negation of a reachability property, BMC presents a series of proof obligations to the underlying SAT/SMT solver in the form:
$BMC(S, P, k) = I(X@0) \land T(X@0, X@1) \land \ldots \land T(X@k\text{-}1, X@k) \land \neg P(X@k)$
for increasing values of $k$ until a trace is discovered.

\subsection{Variable Ordering Example}

\begin{figure}[bth]
	\centering
	\includegraphics[width=\textwidth]{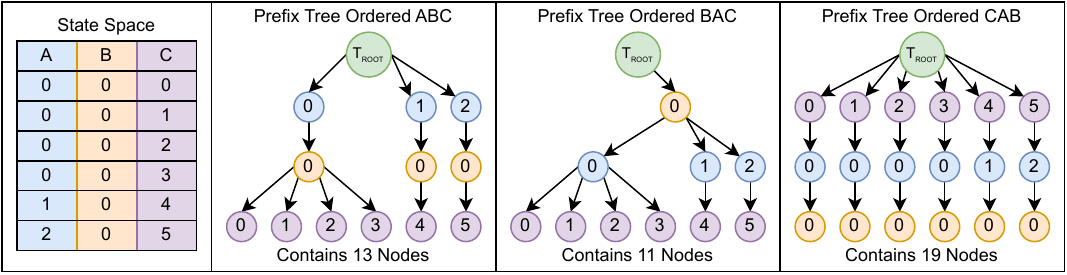}
	\caption{Example Construction of Prefix Tree from State Space}
	\label{fig:ss_ordering}
\end{figure}

Consider the simple example in Figure~\ref{fig:ss_ordering}. Using the default ordering, $A<B<C$, the prefix tree contains 13 nodes. However, by changing the order to $B<A<C$, the state space size reduces to 11 nodes. Using the least optimal ordering, $C<A<B$, the prefix tree consumes nearly double the memory with 19 nodes. This is the crux of the prefix tree's memory efficiency: while a hash map would always store all $18$ variable values, the prefix tree has the potential to use significantly less memory by avoiding duplication. 

If the total number of values for each variable $\species{i}$ is given by $|\species{i}|$, an optimal ordering requires that $\species{i} < \species{j} \iff |\species{i}| < |\species{j}|$.
This creates a dilemma: an optimal variable ordering requires exploring and storing the entire state space, but the variable ordering must be in place for state space exploration and storage. We thus propose the use of BMC-based heuristic discovery as a pre-processing step to assist in bounding variables without the need for full state-space exploration.

\subsection{Variable Bound Calculation}

A probability-agnostic pre-processing step can generate an appropriate variable ordering. This method involves discovering the variable bounds using BMC.  Because these bounds may be too restrictive to include all critical behavior, we do not impose these bounds on the model; rather, we simply use them to create a heuristic. This heuristic is then used to generate a variable ordering.

To begin pre-processing, we construct a dependency graph for the model as described in~\cite{Israelsen2023}. Intuitively, the dependency graph determines transitions that must execute to reach a target state, then it recursively deduces the transitions required to enable the execution of those transitions. 

In the motivating example described in Section~\ref{sec:motivating}, $\mathcal{R}_5$ is the only reaction that increases the count of the target variable $\mathrm{G_{bg}}$, so $\mathcal{R}_5$ is added as a dependency. $\mathcal{R}_5$ is not initially enabled due to a shortage of $\mathrm{RL}$, so $\mathcal{R}_3$ and $\mathcal{R}_8$ are added as dependencies to $\mathcal{R}_5$. Because both $\mathcal{R}_3$ and $\mathcal{R}_8$ are able to execute a sufficient number of times to enable $\mathcal{R}_5$ to produce sufficient $\mathrm{G_{bg}}$ to satisfy the property, the dependency graph terminates with nodes $\mathcal{R}_5$, $\mathcal{R}_3$, and $\mathcal{R}_8$. 

We then encode transitions and initial conditions as a probability-agnostic symbolic transition system.
While this transition system abstracts away the probability information (focusing purely on reachability), the probabilistic property of interest remains the focus of the overall analysis.
This yields a BMC formula of length $k$ (in our case, using \yices~\cite{Yices}). We start with $k=0$ and check the BMC formula using an SMT solver. If the solver returns UNSAT, it means the target is not reachable in $k$ steps, so we increment the value of $k$ by 1 and repeat. When the solver returns SAT, this means that the target is reachable in $k$ steps and that we have found a path from the initial state that reaches the target. The assignment provided by the solver provides one trace. To find bounds, we use the BMC formula in conjunction with the following properties to obtain two sets of bounds for each species or variable. 

The first set of bounds represents the \textit{tightest} bounds, or the most restrictive bounds that allow the execution of \textit{ at least one} \emph{counterexample trace}, that is, a sequence of transitions leading from an initial state to a target state. We iteratively assert that the reachability formula holds in conjunction with an upper and lower bound on each variable until it is impossible to find a trace that satisfies the bounds.

The second set of bounds represents the \textit{loosest} bounds, or the least restrictive bounds that allow \textit{all} counterexample traces to execute. We iteratively assert that the reachability formula holds in disjunction with an upper and lower bound on each variable until no counterexample trace exists outside the bounds. These bounds are used as the primary heuristic to determine a variable ordering for the prefix tree.

\subsection{Bounds-Based Ordering}

For variable \species{i}, the total number of values, which is unknown during this phase, is given by $|\species{i}|$. The tightest lower and tightest upper bounds obtained using the above procedure are given by $\text{Tight}(\species{i}) = [TL(\species{i}),TU(\species{i})]$, respectively. Similarly, the loosest lower and the loosest upper bounds are given by $\text{Loose}(\species{i}) = [LL(\species{i}),LU(\species{i})]$, respectively. For example, if the true values $|\species{i}| = \{3,4,6,19\}$, possible tight and loose bounds may be $\text{Tight}(\species{i}) = [3,6]$ and $\text{Loose}(\species{i}) = [3,19]$. We can then claim that there is \textit{at least one} sequence of transitions leading to a target-satisfying state that causes $\species{i} \in [3,6]$, and \textit{every} sequence of transitions leading to a target-satisfying state produces $\species{i} \in [3,19]$.

To impose an ordering on a pair of variables, we first define the following terms for the variable \species{i}: $L(\species{i}) = LU(\species{i}) - LL(\species{i})$, $T(\species{i}) = TU(\species{i}) - TL(\species{i})$. The following algorithm imposes an order $<$ on a pair of species $(\species{i},\species{j})$:

\begin{enumerate}
	\item If $L(\species{i}) < L(\species{j})$, return $i < j$
	\item If $T(\species{i}) < T(\species{j})$, return $i < j$
	\item Return $\text{random}(j < i, i < j)$
\end{enumerate}

Intuitively, a variable or species with a larger range for its loosest bounds is liable to have more variance than one with a smaller range. If loose bounds are equivalent, the variable or species with a larger range for its tightest bounds is liable to have more variance than the one with a smaller range. If the ranges are equivalent, an order is imposed by a coin flip.

This ordering allows the generation of a memory-efficient prefix tree. While it cannot be guaranteed to be optimal due to an incomplete exploration of the model, we propose that the small amount of time (usually under five minutes) required to generate this ordering using BMC is worth the benefit of improved memory usage in models with very large state spaces.

\subsection{Evaluation of Pre-Processing}

We integrated our C++ benchmarks with \yices~\cite{Yices} via its Python API for BMC-based pre-processing. The pre-processing script implements the techniques described in Section~\ref{sec:preProcessing}, then uses the library \texttt{pybind}~\cite{pybind11} to communicate a variable ordering to the C++ benchmarking program.
 
During exploration of \(10^8\) steps in the modified yeast polarization model (see Section~\ref{sec:motivating}), we note a memory reduction of $10.2$ GB (about 55\%) when employing the pre-processing step in conjunction with guided state space exploration, as opposed to utilizing hash maps. Additionally, we observe a savings of $287$ MB from the pre-processing step when compared to the prefix tree method with random variable ordering and random state space exploration.

The pre-processing step incurs a consistent average time cost of approximately five minutes, independent of the state space size. We suggest that this time investment is justified when analyzing models where memory constraints are a significant concern, especially considering that state space construction for these models can require several hours.

Although the memory savings achieved through the pre-processing step may not be as substantial when compared to random variable ordering as they are against hash maps, we assert that this approach can provide confidence to users that their memory savings will significantly outperform the worst-case scenario. For state spaces that exceed current memory limitations and require hours to days for CTMC transient analysis, we advocate that dedicating a brief period to pre-processing in order to identify an efficient ordering represents a valuable trade-off for very large state spaces where memory efficiency is critical.

%% file: sections/conclusion.tex
\section{Conclusion}
\label{sec:conclusion}

In this work, we have proposed the novel application of a prefix tree data structure to improve memory consumption for large, highly-concurrent CTMCs, including CRNs and other VASes. Prefix trees significantly reduce memory consumption while maintaining constant lookup and insertion times compared to the hash maps employed by cutting-edge tools.
Theoretical complexity analysis indicates that prefix trees are competitive to hash maps, with added potential for memory savings. Benchmarks on realistic, complex CRN models corroborate this analysis, often showing a memory improvement of more than 50\%. 
We further introduce a BMC-based pre-processing step that provides a heuristic to achieve additional memory savings, corroborated by preliminary evaluations.
The adoption of prefix trees, along with the proposed pre-processing method, can lead to significant improvements in memory efficiency for state storage in CTMCs.

Future work includes the exploration of additional heuristics and methods for variable ordering in the prefix tree, as well as integration of the prefix tree data structure into existing PMC tools.